# Lee-Huang-Yang effects in the ultracold mixture of $^{23}$Na and $^{87}$Rb with attractive interspecies interactions

Zhichao Guo,[1] Fan Jia,[1] Lintao Li,[1] Yinfeng Ma,[2] Jeremy M. Hutson,[3] Xiaoling Cui,[2] and Dajun Wang[1,*]

[1]*Department of Physics, The Chinese University of Hong Kong, Hong Kong, China*
[2]*Beijing National Laboratory for Condensed Matter Physics, Institute of Physics,
Chinese Academy of Sciences, Beijing 100190, China*
[3]*Joint Quantum Centre (JQC) Durham-Newcastle, Department of Chemistry, Durham University,
South Road, Durham DH1 3LE, United Kingdom*

The beyond-mean-field Lee-Huang-Yang (LHY) correction is ubiquitous in dilute ultracold quantum gases. However, its effects are often elusive due to the typically much larger influence of the mean-field (MF) energy. In this work, we study an ultracold mixture of $^{23}$Na and $^{87}$Rb with tunable attractive interspecies interactions. The LHY effects manifest in the formation of self-bound quantum liquid droplets and the expansion dynamics of the gas-phase sample. A liquid-to-gas-phase diagram is obtained by measuring the critical atom numbers below which the self-bound behavior disappears. In stark contrast to trapped gas-phase condensates, the gas-phase mixture formed following the liquid-to-gas-phase transition shows an anomalous expansion featuring a larger release energy for increasing MF attractions.

Bose-Einstein condensate (BEC) of dilute atomic gases is a very natural platform for testing the rich physics of interacting Bose gases. While the behavior of BECs is typically dictated by the mean-field (MF) interaction, effects caused by the beyond-MF Lee-Huang-Yang (LHY) correction $E_{\text{LHY}}$ [1] are of great interest due to their deep connections with many-body correlation and quantum fluctuations. With the help of magnetic Feshbach resonances to control the scattering length, the influences of the LHY correction have been studied in various ultracold systems [2–6]. However, typically LHY effects are clearly visible only in the strongly interacting regime with large scattering lengths when $E_{\text{LHY}}$ becomes more or less comparable to MF energy $E_{\text{MF}}$.

The situation can be very different in a double BEC with repulsive intraspecies interactions and a tunable attractive interspecies interaction. In such a system, the total $E_{\text{MF}}$ of the system can be largely canceled or even become negative while the positive $E_{\text{LHY}}$ is nearly intact. As $E_{\text{LHY}}$ is now dominant, its effects become pronounced even though the system is still in the weakly interacting regime [7,8]. At $E_{\text{MF}}$ near zero, the double BEC forms the so-called LHY fluid, which has been studied recently [8,9]. For negative $E_{\text{MF}}$, the system can enter the quantum liquid droplet phase with a signature self-bound behavior in free space [10–13]. Quantum droplets have also been observed in some single-species ultracold lanthanoid systems in which $E_{\text{MF}}$ can be balanced by the dipolar interaction [14–25].

In this work, we study in two different ways the LHY effects in a double BEC of $^{23}$Na and $^{87}$Rb atoms with tunable attractive interspecies interactions. First, we observe the heteronuclear quantum liquid droplet in free space with more than $10^4$ atoms when the interspecies Na-Rb scattering length is tuned into the MF collapse regime. Under optimized conditions, a low-number-density droplet with lifetime exceeding the observation time is observed. We also investigate the liquid-to-gas-phase transition and obtain the critical atom numbers at the phase boundary. Second, we measure the release energies of two types of gas-phase mixtures, the pure in-trap gas and the gas formed after a droplet crosses the liquid-to-gas transition, and observe their opposite dependence on the interaction strength. Coupled with calculations based on extended Gross-Pitaevskii equations (eGPEs), our results confirm the crucial contribution of $E_{\text{LHY}}$ and its effects in stabilizing the heteronuclear double BEC far into the MF collapse region.

Our experiment starts from an optically trapped double BEC of $^{23}$Na and $^{87}$Rb (denoted as species 1 and 2 hereafter, respectively) both prepared in their lowest-energy hyperfine Zeeman level $|F = 1, m_F = 1\rangle$ [26]. To reveal the LHY effects, we focus on the near-zero MF region with $\delta g = g_{12} + \sqrt{g_{11}g_{22}} \approx 0$. Here $g_{ij} = 2\pi\hbar^2 a_{ij}/M_{ij}$ are the two-body interaction constants, with $a_{ij}$ the scattering lengths and $M_{ij}$ the reduced masses. To reach this regime, we use the Na-Rb Feshbach resonance at $B_0 = 347.648$ G to tune the Na-Rb scattering length. The scattering length is represented as $a_{12} = a_{\text{bg}}(1 - \Delta/(B - B_0))$ [upper panel of Fig. 1(a)], with background scattering length $a_{\text{bg}} = 76.33a_0$ and resonance width $\Delta = 4.255$ G [27,28]. Near $B_0$ the intraspecies scattering

*djwang@cuhk.edu.hk

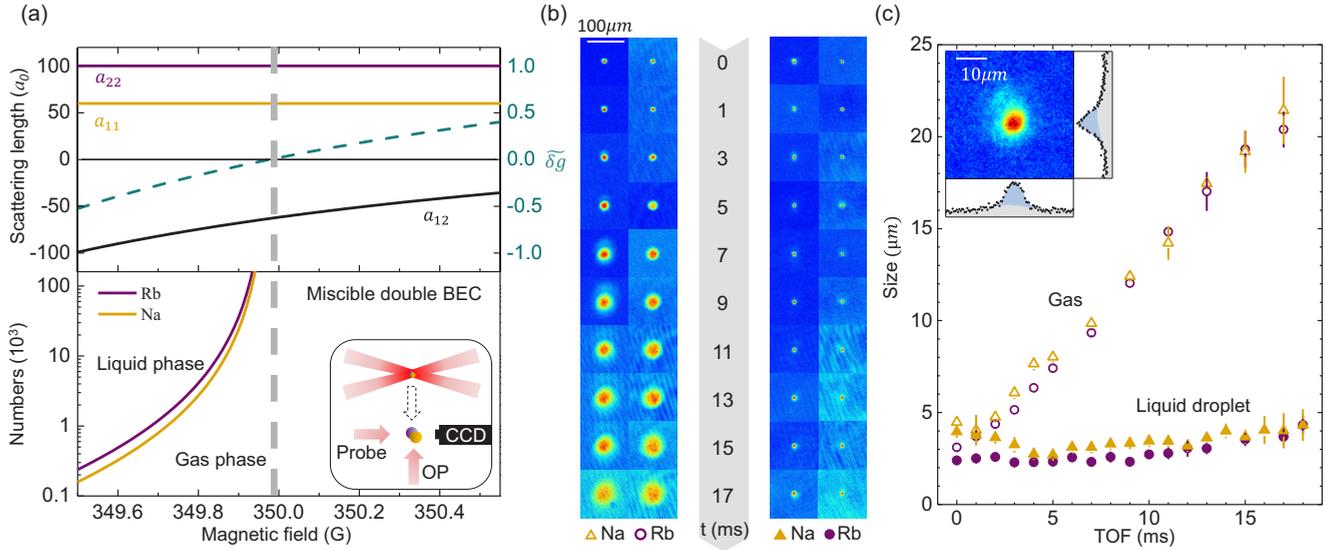

FIG. 1. Creation and detection of Na-Rb mixtures for investigating LHY effects. (a) In the upper panel, the purple, yellow, and black solid lines are scattering lengths between Na-Na($a_{11}$), Rb-Rb($a_{22}$), and Na-Rb($a_{12}$), respectively. The dashed green curve is $\widetilde{\delta g}$. The dashed vertical line indicates the magnetic field for $\widetilde{\delta g} = 0$. In the lower panel, the phase diagrams at different atomic numbers and magnetic fields are shown. Inset: Schematic detection method of the droplet in free space. (b) Images of the mixture during the TOF expansion in the gas phase at 350.451 G (left) and the droplet phase at 349.849 G (right). (c) Cloud sizes obtained from (b) reveal the very different behaviors of the two phases. Data points for $^{87}$Rb are shown in purple, while those for $^{23}$Na are in yellow. Inset: In the droplet phase, bimodal distributions are observed for the cloud with excess atoms. A double Gaussian fitting is used to extract the droplet size in this case.

lengths $a_{11} = 60.05a_0$ [29,30] and $a_{22} = 100.13a_0$ [30,31] remain almost constant.

For convenience, we define a dimensionless parameter $\widetilde{\delta g} \equiv \delta g/\bar{g}$ to characterize the total $E_{\text{MF}}$, with $\bar{g} = (g_{11} + g_{22})/2$. For the current system, the critical magnetic field for $\widetilde{\delta g} = 0$ (and also $\delta g = 0$) is $B_c = 349.978$ G, which corresponds to a critical interspecies scattering length $a_{12}^c = -63.1a_0$. As shown in Fig. 1(a) (lower panel), for each value of $\widetilde{\delta g} < 0$, the double BEC will undergo a transition from gas to droplet phase as the atom numbers are increased. With our atom numbers around $10^4$ for both species, the theory predicts an observation window for the droplet phase for $\widetilde{\delta g}$ from $-0.061$ to $-0.189$ (349.910 G to 349.780 G). For $\widetilde{\delta g}$ from 0 to $-0.061$, droplet formation is still possible but atom numbers much larger than currently available are needed.

In the first experiment, we study the Na-Rb quantum droplet in free space by releasing the sample from the optical trap. We probe the system during the time of flight (TOF), as depicted in the inset of Fig. 1(a). To probe the atoms in situ at each TOF, we implement a two-species high-field detection protocol which consists of a partial optical pumping process followed by standard absorption imaging on the cycling transitions [32]. This partial imaging method is necessary as the droplets have very high optical depth and are difficult to detect directly [11,33]. Both the partial optical pumping and the absorption imaging are calibrated using standard methods [34,35]. The measured resolutions ($1/\sqrt{e}$ half Gaussian widths) of our imaging system are 0.6 and 0.8 $\mu$m for $^{23}$Na at 589 nm and $^{87}$Rb at 780 nm, respectively, which are just enough to resolve the droplet. To maintain the imaging resolutions, we mount the objective and the probe light optics on high-precision vertical translational stages to keep the optical axis of the imaging system always aligned with the droplet.

While shutting down the optical trap provides a direct way of probing the droplet in free space, the nonadiabatic nature of the process can cause problems. Due to the confinement, the in-trap sample is smaller in size than the free-space droplet in its ground state, while the total energy of the system is much larger. In the worst scenario, the energy of the initial state on release from the trap is larger then the energy barrier of the expansion and a stable droplet can never be formed [30]. To mitigate this problem, the crossed optical dipole trap is configured in a nearly spherical shape with measured trap oscillation frequencies of 78 Hz (86 Hz) for $^{87}$Rb ($^{23}$Na). This is in accordance with the understanding that a quantum droplet with isotropic short-range interactions should be spherical in free space. In addition, a carefully designed magnetic-field control sequence is used to improve the mode matching [30]. These steps allow us to observe droplet formation reliably for $\widetilde{\delta g}$ from $-0.094$ to $-0.189$. However, for $\widetilde{\delta g}$ from $-0.061$ to $-0.094$, due to the smaller binding energy, droplets can be observed only by a more sophisticated mode-matching method, aided by fast magnetic-field quenching at the instant of releasing the samples from the trap [30]. For this method to work, the starting magnetic field must be selected carefully so that the size of the in-trap sample matches that of the free-space droplet at the end of the magnetic-field quenching.

Figure 1(b) shows a side-by-side comparison of the $^{23}$Na and $^{87}$Rb clouds in the gas phase and the droplet phase following the TOF. The signature self-bound behavior of the droplet can be observed clearly by comparing its dramatically different TOF expansion with that in the gas phase. For $\widetilde{\delta g} = -0.119$ (349.849 G), where the droplet phase is expected, the sizes of both clouds stay nearly the same during the TOF. For $\widetilde{\delta g} = +0.344$ (350.451 G), by contrast, the system remains in the gas phase, and the sizes increase steadily with TOF.

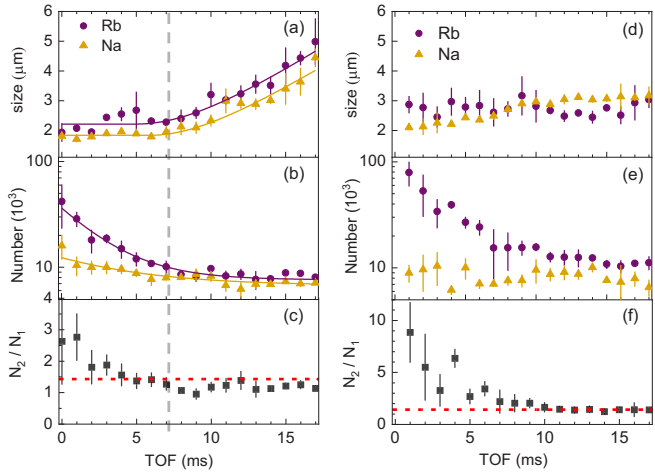

FIG. 2. Evolution of the Na-Rb droplet in free space. Panels (a), (b), and (c) show the size, atom numbers, and the $^{87}$Rb to $^{23}$Na number ratio at $\widetilde{\delta g} = -0.116$ (349.852 G) following the TOF. The droplet is created by directly releasing the sample from the trap. The vertical bar marks the approximate time of the liquid-to-gas-phase transition. Panels (d), (e), and (f) show the evolution for the droplet created with the additional magnetic-field quenching for $\widetilde{\delta g} = -0.089$ (349.880 G). The red dashed lines in (c) and (f) mark the theoretical number ratio $N_2/N_1 = 1.51$.

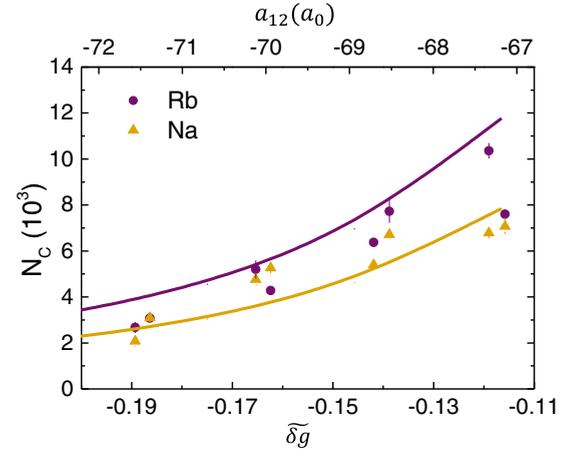

FIG. 3. Critical atom numbers $N_c$ for the liquid-to-gas-phase transition as a function of $\widetilde{\delta g}$. The solid lines are calculated from coupled eGPEs. The color codes are purple for $^{87}$Rb and yellow for $^{23}$Na.

To extract quantitative information, we fit the images with 2D Gaussian functions. In the droplet, the size should be the same for the two species; their densities and thus numbers should follow the ratio $N_2/N_1 = \sqrt{g_{11}/g_{22}} = 1.51$ [7]. When this ratio is not maintained, the droplet part shows up as a dense central peak, and the excess atoms of one species appear as a much larger-sized and expanding gaseous background surrounding the droplet [inset of Fig. 1(c)]. For this kind of "bimodal" distribution, the droplet parameters are extracted with a double 2D Gaussian fitting. Figure 1(c) shows the starkly different expansion behaviors in the droplet and the gas phases obtained from the images in Fig. 1(b).

For the droplets formed by direct trap release from $\widetilde{\delta g} = -0.094$ to $-0.189$, we observe two distinctively different dynamics in the evolution of the $^{23}$Na and $^{87}$Rb samples during the TOF. As an example, Figs. 2(a)–2(c) show the sizes, atom numbers, and number ratio of the droplet sample for $\widetilde{\delta g} = -0.116$ (349.852 G). During the first 10 ms, the $^{23}$Na and $^{87}$Rb sizes increase very little, while atom losses are observed for both species. Bimodal distributions are observed from 0 to 5 ms as the initial number ratio is far from 1.51. After 5 ms expansion, the gaseous atoms surrounding the droplet are already too dilute to be detected.

After about 10 ms, the sample sizes start to increase while the atom numbers stay nearly constant. This is consistent with a phase transition from droplet to gas when the atom numbers are reduced to below the critical values [see Fig. 1(a)]. We fit the size evolution empirically with $\sigma(t > t_0) = \sqrt{\sigma_0^2 + v^2(t-t_0)^2}$ and $\sigma(t < t_0) = \sigma_0$ and obtain a lifetime in the droplet phase of about $t_0 = 7$ ms. Here $v$ is the expansion velocity of the gas-phase sample. Similarly, as illustrated in Fig. 2(b), the critical atom numbers for the phase transition can be obtained by fitting the $^{23}$Na and $^{87}$Rb number evolution data with an exponential decay function with the critical number as the offset.

For several values of $\widetilde{\delta g}$ between $-0.061$ and $-0.094$, we have created longer-lived droplets by the magnetic-field quenching method. As shown in Figs. 2(d)–2(f), during the accessible TOF, the droplet size stays nearly the same although number losses are still observed. After 10 ms, the number ratio becomes close to 1.51 and the number loss slows down significantly. As the lifetime is longer than the usable TOF of 18 ms, we have not been able to measure the real lifetime and the critical numbers in this range of $\widetilde{\delta g}$. More detailed investigations, e.g., by levitating the droplet, are warranted in the near future.

Figure 3 shows the critical $^{23}$Na and $^{87}$Rb atom numbers for $\widetilde{\delta g}$ from $-0.094$ to $-0.189$. Although the range of $\widetilde{\delta g}$ is small, a fourfold change in the measured critical numbers is observed. The critical numbers calculated using coupled eGPEs with the LHY term included (solid curves) matches the measurements very well.

A major cause of the atom losses is three-body recombination as a result of the high number densities in the droplet [10–12,30]. For droplets with $\widetilde{\delta g}$ from $-0.094$ to $-0.189$, another possible loss mechanism is self-evaporation due to the imperfect matching between in-trap and free-space modes [36]. The short lifetimes in this region are the combined result of these loss mechanisms. In contrast, the droplet at $\widetilde{\delta g} = -0.089$ [see Figs. 2(d)–2(f)] has relatively lower number densities and is formed with nearly optimal mode matching and is thus longer lived. Nevertheless, for both cases, the stabilized number ratios after 10 ms TOF are very close to 1.51 despite the different creation procedures and the large initial number mismatches [Figs. 2(c) and 2(f)].

We now turn to the gas-phase double BEC and release the sample directly from the trap (without the additional magnetic-field quenching) for $\widetilde{\delta g}$ from $+0.344$ to $-0.094$. To characterize the LHY effect, we measure the release energy $E_{\rm rel} = E_{\rm kin} + E_{\rm int}$ from the TOF expansion [30,37,38]. Here $E_{\rm kin}$ is the quantum pressure and $E_{\rm int} = E_{\rm MF} + E_{\rm LHY}$ is the total interaction energy. When $E_{\rm MF}$ is tuned to zero, $E_{\rm LHY}$ becomes the only interaction term in $E_{\rm rel}$. For negative $E_{\rm MF}$, the fact that the sample expands rather than collapses is also a direct manifestation of the LHY effect.

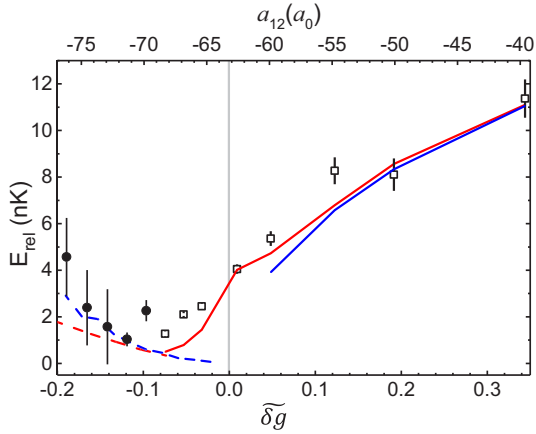

FIG. 4. The release energy as a function of $\widetilde{\delta g}$. The open squares represent $E_{\rm rel}$ for gas-phase mixtures released from the trap directly, while the solid circles are $E_{\rm rel}$ for the gas samples formed following the liquid-to-gas-phase transition. The red solid curve and the red dashed curve show $E_{\rm rel}$ for these two cases, calculated with eGPEs. The magenta short dashed curve is $E_{\rm rel}$ for the second case from variational theory. The blue solid curve shows the GPE calculation without the LHY term for comparison.

We prepare the trapped double BEC and measure $E_{\rm rel}$ from the TOF expansion velocities of the two species [30,37] for several different values of $\widetilde{\delta g}$. The resulting $E_{\rm rel}$ is shown as open squares in Fig. 4. In general, $E_{\rm rel}$ depends on $\widetilde{\delta g}$, $N_1$, and $N_2$ and the density overlap between the two species. However, in this range of $\widetilde{\delta g}$, the densities of the double BEC are still rather high even in the gas phase, and significant three-body losses are observed during the TOF. Thus, it is difficult to maintain constant $N_1$ and $N_2$ during the TOF for different $\widetilde{\delta g}$, even though great efforts are made to keep the initial atom numbers in the trap fixed. Nevertheless, since $E_{\rm rel}$ is a measure of the energy per atom, it is not affected severely by the loss.

We calculate $E_{\rm rel}$ with the eGPEs for the corresponding atom numbers at the end of the TOF and find good agreement for $\widetilde{\delta g} \gtrsim 0$, corresponding to $E_{\rm MF} \gg E_{\rm LHY}$ or $E_{\rm MF} \approx 0$ (red solid curve in Fig. 4). We note that, without the LHY term, the calculation fails for $\widetilde{\delta g} < 0$ due to collapse of the double BEC (blue solid curve in Fig. 4). At $\widetilde{\delta g} \approx 0$, where $E_{\rm MF}$ almost cancels, the energy difference between the two calculations is about 1 nK. This is essentially the magnitude of $E_{\rm LHY}$, if we assume $E_{\rm kin}$ does not change dramatically between the two cases.

The gas-phase mixture formed following the liquid-to-gas-phase transition is also of great interest. As shown in Fig. 2(a), its expansion velocity can be measured and converted into another $E_{\rm rel}$ even though there is no trap involved. Following the number losses, the total energy of the system changes from negative in the droplet phase to positive in the metastable and eventually gas phase. $E_{\rm rel}$ here is thus just the sum of $E_{\rm kin}$, $E_{\rm MF}$, and $E_{\rm LHY}$ at the phase-transition point [30]. The measured $E_{\rm rel}$ is shown in Fig. 4 as solid circles for $\widetilde{\delta g}$ from $-0.094$ to $-0.189$. It has the opposite dependence on $\widetilde{\delta g}$ from that observed for the gas-phase mixture released from the trap. This is similar to previous observations on the expansion dynamics of the erbium dipolar system [17].

This behavior can be qualitatively understood from the competition between $E_{\rm LHY}$, $E_{\rm kin}$, and $E_{\rm MF}$ for the droplet near the phase-transition point. As $\widetilde{\delta g}$ becomes more negative, the critical atom numbers at the phase-transition point become smaller, and the sample also shrinks. The positive $E_{\rm LHY}$ and $E_{\rm kin}$ terms increase faster than the negative $E_{\rm MF}$ term decreases. The measured $E_{\rm rel}$ thus increases for more negative $\widetilde{\delta g}$. We note that, without $E_{\rm LHY}$, the magnitude of $E_{\rm MF}$ is larger than $E_{\rm kin}$ and the system collapses. This picture is confirmed semiquantitatively by the eGPE calculation, shown by the red dashed curve in Fig. 4. As a comparison, a variational calculation (magenta short dashed curve) using the experimentally measured atom numbers and the Gaussian ansatz is also shown; this actually agrees better with the experimental data. The deviation between the experiment and the eGPE calculation may be due to the inaccuracy in determining the transition point.

In conclusion, we have studied the heteronuclear BEC mixture of $^{87}$Rb and $^{23}$Na atoms with attractive intersperses interactions under conditions where the role of the LHY correction is amplified. The signatures of the LHY correction manifest clearly in both the self-bound behavior of the free-space quantum liquid droplet and the expansion of the gas-phase mixtures. Although the observation time is limited by the detection method, our result shows that heteronuclear droplets can be long lived. This will allow quantitative studies on the fundamental properties of the quantum droplet, such as the existence of a possible Bose pairing mechanism [39,40] with mass-imbalanced constituents. Another interesting topic is quantum droplets in lower dimensions and at the crossover between different dimensionalities [41–43]. The $^{23}$Na and $^{87}$Rb system here also holds great promise in creating large quantum liquid droplets with more than $10^6$ atoms [44,45]. In this so far unreached regime, the droplet features a large core with constant density, and various peculiar collective excitation may be investigated [7].

We thank valuable discussions with D.S. Petrov, L. Tarruell, L. Tanzi, Zeng-Qiang Yu, and Hui Hu. This work was supported by the Hong Kong RGC General Research Fund (Grants 14301620 and 14303317) and the Collaborative Research Fund C6005-17G. J.M.H. is supported by the U.K. Engineering and Physical Sciences Research Council (EPSRC) Grant No. EP/P01058X/1. Y.M. and X.C. are supported by the National Key Research and Development Program of China (2018YFA0307600, 2016YFA0300603) and the National Natural Science Foundation of China (12074419).

## APPENDIX

### 1. The intraspecies scattering lengths

Near 350 G, there are no intraspecies Feshbach resonances for either $^{23}$Na or $^{87}$Rb atoms in their $|1, 1\rangle$ hyperfine Zeeman states. However, the $a_{11}$ and $a_{22}$ in this region still need to be treated carefully. To this end, we calculate the scattering

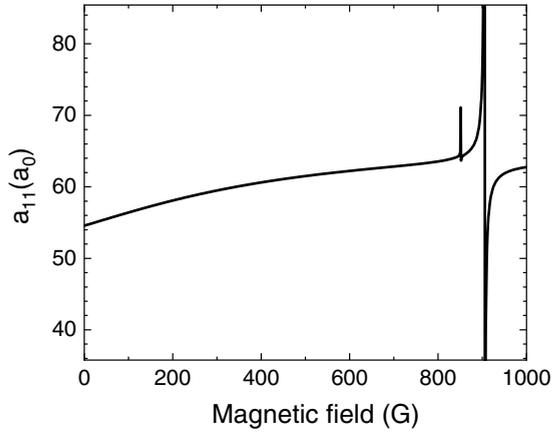

FIG. 5. The intraspecies scattering length for $^{23}$Na atoms in the $|1, 1\rangle$ state. The two resonances at high magnetic fields around 900 G [29] are far away from the current magnetic field of interest near 350 G. However, $a_{11}$ still varies smoothly from $54.5a_0$ at 0 G to $60.05a_0$ for 350 G.

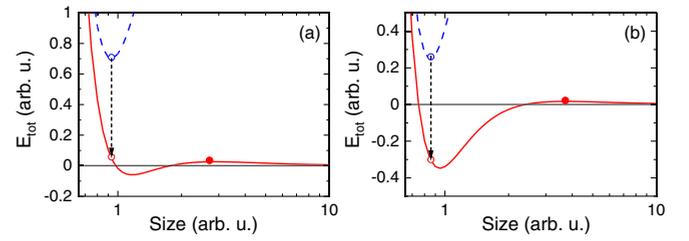

FIG. 6. Droplet formation by directly releasing the trapped sample. The blue dashed curves represent $E_{\text{tot}} = E_{\text{kin}} + E_{\text{pot}} + E_{\text{int}}$ of the trapped double BEC, and the red solid curves are $E_{\text{tot}}$ in free space (i.e., $E_{\text{kin}} + E_{\text{int}}$). (a) For very small $|\widetilde{\delta g}|$, the free-space $E_{\text{tot}}$ (the red open circle) is higher than the barrier height on the large size side (the red solid dot). In this case, the sample will keep expanding and never forms the droplet. (b) For $|\widetilde{\delta g}|$ large enough, the sample does not have enough energy and will be confined near the energy minimum to form the droplet.

lengths with the MOLSCAT package [46,47] using the molecular potentials in Ref. [29] for $^{23}$Na and in Ref. [31] for $^{87}$Rb. For $^{87}$Rb, the scattering length $a_{22}$ near 350 G stays nearly the same as the well-calibrated low-field value of $100.13a_0$. On the other hand, for $^{23}$Na, the scattering length $a_{11}$ varies smoothly as the magnetic field is changed from 0 to 350 G, as shown in Fig. 5. Near the experimentally used magnetic field for the current experiment, $a_{11}$ is $60.05a_0$. This is rather different from the low-field value of $54.5a_0$.

The variation of $a_{11}$ far away from the Feshbach resonance is due to the change of the singlet/triplet character of the $^{23}$Na atomic states. This is the result of the competition between the hyperfine interaction and the Zeeman effect. Since the difference between the singlet and triplet $^{23}$Na - $^{23}$Na scattering lengths is large [29], $a_{11}$ will be modified by the singlet/triplet character variation. For $^{87}$Rb, since the singlet and triplet scattering lengths are very similar [31], $a_{22}$ is very insensitive to magnetic fields unless there is a Feshbach resonance.

#### 2. Magnetic field gradient compensation

An important technical issue for observing the free-space droplet is the spatial varying magnetic field experienced by the droplet during the TOF. Due to the different magnetic dipole moments of $^{23}$Na ($\mu_{b1} = 1.084$ MHz/G) and $^{87}$Rb ($\mu_{b2} = 0.837$ MHz/G) near $B_0$, a moderate gradient and curvature may generate a differential force large enough to tear the droplet apart. With the pair of Feshbach coils only, the measured magnetic field curvature near 347 G is 1.4 G/cm$^2$ and the averaged magnetic field gradient is about 0.6 G/cm. Experimentally, we observe the $^{23}$Na and $^{87}$Rb atomic clouds separating from each other following the TOF. To mitigate this problem, we placed another coil, with its current dynamically controlled following the TOF, on top of the Feshbach coil. This simple improvement reduces the magnetic field gradient to 0.11 G/cm and the magnetic field variation during the 18 ms TOF from $\pm 10$ mG to $\pm 3$ mG, which corresponds to a $a_{12}$ change of less than $0.26\, a_0$.

In the eGPEs, the magnetic field gradient is included as opposite direction linear potentials to the two species in the center-of-mass coordinate. According to the simulation, this small magnetic field gradient actually doubles the critical atom numbers.

#### 3. In-trap and free-space mode matching

During the release of the double BEC sample from the optical trap, the trap energy suddenly is removed suddenly, but the size of the sample has no time to adapt to the new value. As shown in Fig. 6, in general the in-trap size is smaller than the size of the ground-state free-space droplet at the energy minimum. The sample size will thus start to expand after being released from the trap and its total energy (per particle) $E_{\text{tot}} = E_{\text{kin}} + E_{\text{int}}$ will also start to evolve following the red curves. For the cases with very small $|\widetilde{\delta g}|$ [Fig. 6(a)], $E_{\text{tot}}$ is higher than the energy barrier at the right-hand side. In this case, the sample will just keep expanding after crossing the maximum point and never forms the droplet, while in Fig. 6(b), with $|\widetilde{\delta g}|$ large enough, the negative $E_{\text{tot}}$ of the released sample is not enough to overcome the energy barrier and the droplet can be formed, but its size will undergo a small-amplitude oscillation.

As described in the main text, to improve the mode matching, we used a nearly spherical shape trap for matching the sphere shape of the quantum droplet. In addition, as the in-trap size of the double BEC also depends strongly on $\widetilde{\delta g}$, the magnetic field is controlled in a carefully designed sequence. After the double BEC is created, the magnetic field is first ramped up quickly across the Feshbach resonance to 358.4 G where $a_{12}$ has a small positive value of $46.1a_0$. After 350 ms for the system to stabilize, the magnetic field is ramped down to 350.451 G in 130 ms. This is just 0.424 G above $B_c$ and the corresponding $a_{12}$ is $-39.5a_0$ ($\widetilde{\delta g} = +0.344$). Here the double BEC is miscible and the attractive interaction increases the spatial overlap of the two condensates and also increases the in-trap densities substantially. Both effects significantly improve the mode matching and facilitate the free-space droplet formation. At this point, rapid atom losses are already observed; thus, the magnetic field holds here for only 10 ms.

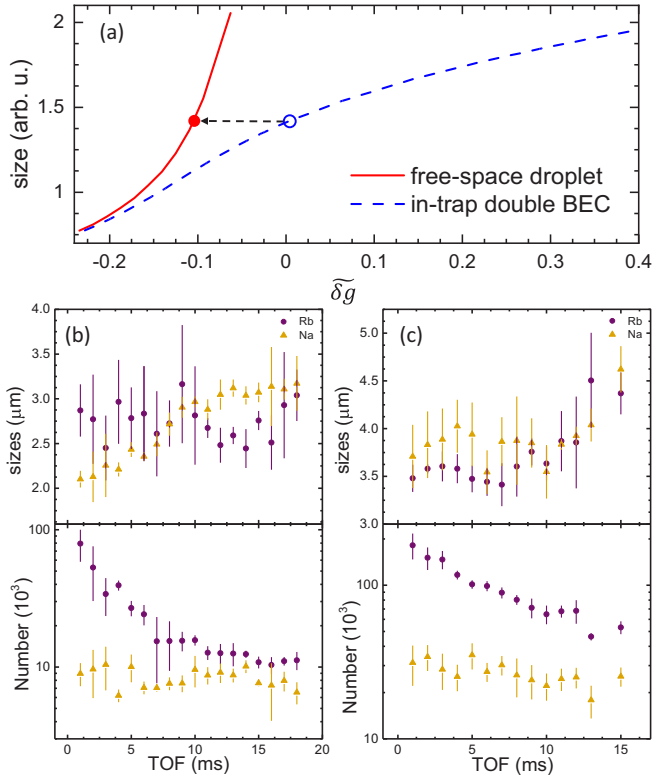

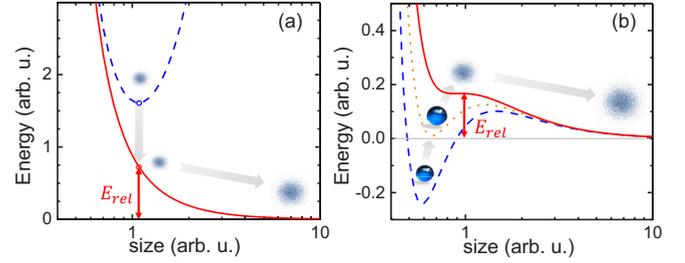

FIG. 7. Producing long-lived droplets by magnetic field quench. (a) For some weakly bound droplets, it is possible to find a $\widetilde{\delta g}$ value where the in-trap gas-phase sample size (blue dashed curve) is the same as the size of the free-space droplet (red solid curve). Quenching $\widetilde{\delta g}$ at the instant of releasing the sample, as depicted by the black dashed arrow, will lead to a near perfect mode matching. Panels (b) and (c) show the experimentally measured droplet signals for $\widetilde{\delta g} = -0.089$ (349.880 G) and $\widetilde{\delta g} = -0.061$ (349.910 G) created with this method. At these magnetic fields, no droplet can be observed without the quench as they fall in the situation as described in Fig. 6(a).

FIG. 8. The release energies. (a) Origin of $E_{\rm rel}$ for the gas released from the trap. The blue dashed curve is the in-trap total energy $E_{\rm tot} = E_{\rm pot} + E_{\rm kin} + E_{\rm int}$, and the red solid curve is $E_{\rm tot}$ without $E_{\rm pot}$ when the sample is suddenly released from the trap. (b) $E_{\rm rel}$ for the case following the liquid-to-gas-phase transition. The blue and orange dashed curves are $E_{\rm tot}$ of the stable and meta-stable droplets, respectively. The red solid curve is $E_{\rm tot}$ just on the phase-transition point, where the local minimum disappears. For both cases, following the size increase, the quantum pressure $E_{\rm kin}$ and the interaction energy $E_{\rm int}$ are converted to the kinetic energy of the atoms.

Afterward, the magnetic field is ramped to a target value within 5 ms before the droplet is detected in free space.

As depicted in Fig. 6(b), although the droplet can be observed after these improvements for $|\widetilde{\delta g}|$ large enough, the mode matching is not perfect. For $\widetilde{\delta g}$ from 0 to $-0.094$, with the current atom numbers, the droplet cannot be observed by simply releasing the double BEC from the trap. As shown in Fig. 7(a), even better mode matching can be achieved by finding a $\widetilde{\delta g}$ value where the in-trap size matches with the target droplet size and quenches $\widetilde{\delta g}$ at the same time when the trap is shut off. To achieve the fast magnetic-field quenching necessary for this method, we added a small magnetic coil that is capable of covering the magnetic field range in less than 10 $\mu$s. Figures 7(b) and 7(c) show two droplets formed with this method. For case (b) with $\widetilde{\delta g} = -0.089$, the droplet lifetime is longer than the available observation time. For case (c) with $\widetilde{\delta g} = -0.061$, the liquid-to-gas-phase transition is observed after about 7 ms. This short lifetime is probably limited by the atom numbers since at this $\widetilde{\delta g}$, the critical atom numbers are much larger.

### 4. Three-body loss

As discussed in the main text, the major losses of atom number are from a three-body loss as a result of the high density of the droplet sample. We observe a faster loss of $^{87}$Rb than that of $^{23}$Na at almost all magnetic fields. We believe this is due to the larger three-body loss rate constant ($7.2 \times 10^{-30}$ cm$^6$/s) of the $^{87}$Rb BEC than that of the $^{23}$Na BEC ($1.1 \times 10^{-30}$ cm$^6$/s) [48,49].

While the three-body loss rate constants for condensed $^{87}$Rb and $^{23}$Na atoms are well known [48,49], those between $^{23}$Na and $^{87}$Rb with the weak interspecies interaction near $B_c$ are unknown. But from our previous few-body investigations [50], no heteronuclear Efimov resonances are expected near this region; thus, we can assume that the heteronuclear three-body loss should have a similar rate as single species $^{87}$Rb or $^{23}$Na samples.

### 5. Release energies

For the double BEC system studied here, the average release energy is defined as $E_{\rm rel} = (\frac{1}{2}m_1v_1^2N_1 + \frac{1}{2}m_2v_2^2N_2)/(N_1 + N_2)$, where $v_i$ is the expansion velocity. Figure 8 illustrates schematically the different origins of $E_{\rm rel}$ for the in-trap gas and the gas formed from the droplet. For the first case [Fig. 8(a)], $E_{\rm rel}$ is dominated by $E_{\rm MF}$ and decreases for smaller $\widetilde{\delta g}$. For the gas from phase-transition case [Fig. 8(b)], the total energy of the sample and thus $E_{\rm rel}$ increases when $\widetilde{\delta g}$ gets more negative.

Experimentally, the RMS sizes are obtained from fitting to the absorption images. Different fitting functions are used for the two cases. For the sample directly released from the trap, the images are fitted with the parabola function in accordance with the Thomas-Fermi approximation for BECs with more than $10^4$ atoms. For the gas mixtures formed following the liquid-to-gas transition, the Gaussian function, which is closer to the small atom number sample profiles, is used.

### 6. Numerical simulation

For the condensate mixture of $^{23}$Na and $^{87}$Rb (denoted as species 1 and 2 hereafter, respectively) considered here, the coupled eGPEs including the LHY correction reads

$$i\hbar \frac{\partial \psi_i}{\partial t} = \left(-\frac{\hbar^2 \nabla^2}{2m_i} + V_i + \mu_i(n_1, n_2)\right)\psi_i, \quad \text{(A1)}$$

with $m_i$ the atomic mass, $n_i$ the number density, and

$$\mu_i = g_{ii}n_i + g_{ij}n_j + \frac{\delta \mathcal{E}_{\text{LHY}}}{\delta n_i}, \quad (i \neq j). \quad \text{(A2)}$$

As introduced in Refs. [7,51], the LHY term for mass-imbalanced system can be described as

$$\mathcal{E}_{\text{LHY}} = \frac{8}{15\pi^2} \frac{\left(m_1^{3/2}(g_{11}n_1)\right)^{5/2}}{\hbar^3} f(\gamma, x, y), \quad \text{(A3)}$$

with $\gamma = m_2/m_1$, $x = g_{12}^2/g_{11}g_{22}$, and $y = g_{22}n_2/g_{11}n_1$. The dimensionless parameter $f(\gamma, x, y)$ represents the renormalized integral with only numerical solution for the mass-imbalanced case.

### 7. Influence of the residue magnetic field gradient

As discussed in the Appendix, Sec. A 2, the magnetic field gradient compensation is not perfect in the current setup. Although small, the residue gradient can in principle generate some influence to the droplet. The gradient $\partial B/\partial y$ is mainly along the vertical (y) direction. In the center-of-mass coordinate, the differential magnetic dipole moments lead to opposite direction linear potentials to the two species,

$$V_i = \left(-\mu_{bi}\frac{\partial B}{\partial y} - a_c m_i\right)y, \quad \text{(A4)}$$

where $a_c$ is the center-of-mass acceleration defined as

$$a_c = \frac{(\mu_{b1}N_1 + \mu_{b1}N_2)}{N_1 m_1 + N_2 m_2} \frac{\partial B}{\partial y}. \quad \text{(A5)}$$

Including $V_i$ in the eGPE, our simulation shows that the critical numbers needed for droplet formation can indeed be increased. The theoretical curve in Fig. 3 does not take this contribution into account. In future experiments, this should be investigated in more detail, for example, by studying the phase transition at various magnetic field gradients. For a better gradient compensation, additional coils will have to be installed.